\documentclass[12pt]{iopart}
\usepackage{iopams}
\usepackage{graphics}
\begin{document}

\title[Time-Dependent Wormhole Solutions of $f(R)$ Theory of Gravity]{Time-Dependent Wormhole Solutions of $f(R)$ Theory of Gravity and Energy Conditions}

\author{H. Saiedi $^{1}$  and  B. Nasr Esfahani  $^{2}$ } \ \\
\address{Department of Physics, University of Isfahan,  Iran.}
\ead{\mailto {$^{1}$hrssaiedi@yahoo.com}, \mailto {$^{2}$ba\_ nasre@sci.ui.ac.ir}}
\begin{abstract}\noindent  In this work, we construct  time-dependent wormhole solutions in the context of $f(R)$ theory of gravity. The background matter is considered to be traceless. By considering specific shape function and power-law expansion exact solutions for $f(R)$ are found. The null and the weak energy conditions (NEC and WEC) are checked for wormhole solutions. It is shown that the matter threading the wormhole spacetimes  with either accelerated expansion or  decelerated expansion satisfies the NEC and WEC. \\

\end{abstract}

\section{Introduction}
Recent astronomical observations have shown that the universe is accelerating at present time [1, 2]. According to this nowadays standard scenario, the universe is flat and undergoing an accelerated  expansion driven by a mysterious  fluid with negative pressure nearly homogeneously distributed and making up to $\sim 70 \ \% $ of the energy content. This exotic component is called dark energy. The dark energy is
gravitationally repulsive and accelerates the expansion of the universe.  \\
There are different proposals for explaining this phenomenon. Some of them are based on assumptions of standard cosmological models, which utilize FRW metric. The simplest possibility for dark energy is a cosmological constant $\Lambda$, introduced by Einstein in 1917. In the early history of modern cosmology the cosmological constant was invoked twice. First by Einstein to obtain static models of the Universe. Next by Bondi and Gold and by Hoyle to resolve an age crisis and to construct a Universe that satisfied the ``Perfect Cosmological Principle," i.e., one that appears the same at all times and places. In both instances the motivating crisis passed and the cosmological constant was put aside. The cosmological constant corresponds to a fluid with a constant equation of state parameter $\omega = -1$ [3, 4].

The other possibility to dark energy related with the modifications of Einstein gravity in such a way, that would give the gravitational alternative to dark
energy. Conceptually, it looks very attractive as then the presence of dark energy is only the consequence of the universe expansion. Examples of such modified gravity
Dark Energy (DE) models are theories  where the Ricci scalar R in the Lagrangian is replaced by a generic function $f(R)$. The usual general relativity is recovered in the limit $f(R)= R$, while completely different results may be obtained for other choices of $f(R)$. Modified $f(R)$ gravity presents very natural unification of
the early-time inflation and late-time acceleration thanks to different role of gravitational terms relevant at small and at large curvature. Earlier interest in $f(R)$ theories was motivated by inflationary scenarios as for instance, in the  Starobinsky model, where $f(R) = R - \Lambda + \alpha R^2$ was considered [5]. In fact, it was shown that the late-time cosmic acceleration can be indeed explained within the context of $f(R)$ gravity [6].  \\
As in general relativity, $f(R)$ gravity theories obtain the field equations by varying the total action for both the field and matter. In the literature, there are two approaches on how to perform the variation. we use the metric (Einstein - Hilbert) variational principle, according to which the action is varied with respect to the metric tensor $g_{\mu\nu}$, and the affine connection coefficients are the Christoffel symbols of $g_{\mu\nu}$. The other one is the metric-affine (Palatini) variational principle, according to which the metric and connection are considered as  geometrically independent quantities, and the action is varied with respect to both of them. the field equations in the metric theories are fourth-order differential equations, while for the Palatini formalism they are second-order.

In this paper,  we extend the analysis of dynamic and spherically symmetric spacetimes considered in the literature, and analyze traversable wormhole solutions in $f(R)$ modified theories of gravity. Traversable wormholes were popularized as a respectable
theoretical possibility by Morris and Thorne in 1988 [7]. The spatial topology of the wormhole is the same as in the black-hole cases, but the throat or the minimal surface is preserved in time, so that observers can pass through it in either direction. In classical general relativity, wormholes are supported by negative-energy matter, i.e., exotic matter, which involves a stress-energy tensor that violates the null energy condition (NEC) [7, 8].
Hochberg and Visser proof in their work that the violation of the NEC is a generic feature of all wormholes, whether they are time-dependent or static [9]. Also, they have shown that for evolving wormholes, the violation of the weak energy condition (WEC) can be avoided, but the NEC is not preserved [10-12]. The nature of the energy condition violations associated with wormholes has led numerous people to try to find ways of evading or minimizing the violations. Several candidates have been proposed in the literature, amongst which we refer to solutions in higher dimensions, for instance, Bhawal and Kar investigated the energy conditions for wormhole solutions in the context of $D$-dimensional Einstein-Gauss-Bonnet theory of gravitation [13].
DeBenedictis and Das have studied wormhole solutions to Einstein's equations with an arbitrary number of time-dependent compact dimensions; for these solutions radial size of energy condition violation is restricted by the extra dimensions [14]. Bronikov and Kim have studied possible wormholes in brane world [15].

In this work, we will investigate time-dependent  wormholes  for the modified $f(R)$ gravity. We will confine the argument to dynamic wormholes in an expanding universe dominated by traceless fluid and will discuss the  energy conditions (NEC and WEC) violation by these solutions.  \\

\section{ Field Equations in $f(R)$ Theory of Gravity}
In this section we shall analyze the modified $f(R)$ gravity action in  four dimensions spacetime. A spherically symmetric metric tensor will be considered. We shall present some geometric properties associated with this corresponding spacetime, and provide the general fields equations obeyed by the components of the metric tensor.

Let us start from the rather general 4-dimensional action [16] \\
\begin{equation}
A=A_g + A_m = \int {\sqrt{-g} \ (f(R) + 2L_m) \ d^4x} \ ,
\end{equation} \ \\
where R is the scaler curvature, $f(R)$  is an arbitrary function in terms of R  and $L_m$ \ is a matter Lagrangian density. The signature is $(- + + +)$ and we are using physical  units $8\pi G = c=1$. ( $G$ is Newton's gravitational constant and $c$ is the speed of light).

By the variation of the action (1) with respect to the metric
$g_{\mu\nu}$, one reaches to the following field equation \\
\begin{equation}
FR_{\mu\nu} - \frac{1}{2}g_{\mu\nu}f -
\nabla_{\mu}\nabla_{\nu}F + g_{\mu\nu}\opensquare F - T_{\mu\nu}^{m} = 0 \ ,
\end{equation} \ \\
which are fourth-order equations. $ F = df/dR $, $R_{\mu\nu}$ is the Ricci tensor and $\mu , \nu=0, 1, 2, 3$. Considering the contraction of equation (2), provides the following relationship \\
\begin{equation}
FR-2f+3\opensquare F = T \ ,
\end{equation} \ \\
which shows that the Ricci scalar is a fully dynamical degree of freedom, and $T = T^{\mu}_{\mu}$ is the trace of the stress-energy tensor.
The trace equation (3) can be used to simplify the field equations and then can be kept as a constraint equation. Thus, substituting the trace equation into equation (2), and re-organizing the terms we end up with the following gravitational field equation \\
\begin{equation}
G_{\mu\nu} = R_{\mu\nu} - \frac{1}{2} g_{\mu\nu} R = {T}^{e}_{\mu\nu} \ ,
\end{equation} \ \\
where the effective stress-energy tensor is given by ${T}^{e}_{\mu\nu} =\tilde{T}_{\mu\nu} + \overline{T}_{\mu\nu}^m$ . The term $\tilde{T}_{\mu\nu}$ is given by
\begin{equation}
\tilde{T}_{\mu\nu} = \frac{1}{F} \left [ \nabla_\mu\nabla_\nu
F -   \frac{1}{4}  g_{\mu\nu} (T+\opensquare F+RF )\right ] \ ,
\end{equation} \ \\
is called the curvature stress-energy tensor, and
\begin{equation}
\overline{T}_{\mu\nu}^m = \frac{ 1}{F} T_{\mu\nu}^m \ .
\end{equation} \ \\
It must be noted that for $f(R)= R$, the curvature stress-energy
tensor identically vanishes and (4) reduces to the standard
second-order Einstein field equations. It is clear that the
curvature stress-energy tensor plays the role of a further source
of gravitation in the field equations.

Now, consider the following spherically symmetric metric \\
\begin{equation}
ds^2 = - e^{2\Phi(r)} dt^2 + S(t)^2 \left[ \frac{dr^2}{1 - \frac{b(r)}{r}
} +r^2 d\theta^2 + r^2 \sin^2 \theta d\phi^2 \right] \ ,
\end{equation} \ \\
which is the general metric for a dynamic wormhole spacetime [17]. $S(t)$ is the scale factor, $\Phi(r)$ and $b(r)$ are arbitrary functions of the radial coordinate, $r$, denoted as the redshift function, and the shape function, respectively. The radial coordinate $r$ is non-monotonic in that it decreases from infinity to a minimum value $r_0$, representing the location of the throat of the wormhole, where $b(r_0)=r_0$, and then it increases from $r_0$ back to infinity. For the wormhole to be traversable, one must demand that there are no horizons present, so that $\Phi(r)$ must be finite everywhere. In the analysis outlined below, we consider that the redshift function is constant, $\Phi'=0$, which simplifies the calculations considerably, and provide interesting exact wormhole solutions.

Relative to the matter content of the wormhole, we impose that the stress-energy tensor that threads the wormhole satisfies the energy conditions. Using an  orthonormal  reference frame with the basis \\
\begin{eqnarray}
\eqalign {\textbf{e}_{\hat{0}} = e^{-\Phi}\textbf{e}_t \ , \\
\textbf{e}_{\hat{1}} = \frac{\textbf{e}_r \sqrt{(1-b/r)}}{S} \ , \\
\textbf{e}_{\hat{2}} = \frac{\textbf{e}_{\theta}}{rS} \ , \\
\textbf{e}_{\hat{3}} = \frac{\textbf{e}_{\phi}}{rS\sin\theta} \ ,}
\end{eqnarray} \ \\
the stress-energy tensor is given by \\
\begin{equation}
T_{\hat{\mu}\hat{\nu}} = diag ( \rho(r), p_r(r), p_t(r), p_t(r) ) \ ,
\end{equation} \ \\
where $\rho$ denotes the energy density, $p_r(r)$ is the radial pressure measured in the direction of $\textbf{e}_{\hat{1}}$, and $p_t(r)$ is the transverse pressure measured in the orthogonal direction to $\textbf{e}_{\hat{1}}$.
It can be shown that in the orthonormal frame (8), the metric (7) becomes \\
\begin{equation}
g_{\hat{\mu}\hat{\nu}} = diag ( -1, 1, 1, 1 ) \ .
\end{equation} \ \\
The curvature scalar for the metric (7) depends
on $r, t$, that is \\
\begin{equation}
R = 6(\dot{H} + 2H^2) + \frac{2b'}{S^2r^2} \ ,
\end{equation} \ \\
where $H=\dot{S}/S$, is the Hubble parameter. The prime denotes a derivative with respect to the radial coordinate, $r$, and the overdot denotes differentiation with respect to time. The effective field equation (4) provides the following relationships \\
\begin{eqnarray}
3H^2 + \frac{b'}{S^2r^2} = \frac{\rho}{F} + \frac{1}{F} \left[ N + \ddot{F} \right] \ , \\ \nonumber \\
\fl - 2\dot{H} -3H^2 - \frac{b}{S^2r^3} =  \frac{p_r}{F} + \frac{1}{F} \left[ -N -H\dot{F} + \frac{(r-b)F''}{S^2r}  -  \frac{(b'r-b)F'}{2S^2r^2} \right]  \ , \\ \nonumber \\
- 2\dot{H} -3H^2 - \frac{b'r-b}{2S^2r^3} = \frac{p_t}{F}  +  \frac{1}{F} \left[ -N - H\dot{F} + \frac{(r-b)F'}{S^2r^2} \right] \ ,
\end{eqnarray} \ \\
where $F=F(r,t)$, $\dot{F}=(\partial/\partial t)F(r,t)$, $\ddot{F}=(\partial/\partial t)^2F(r,t)$, $F' = (\partial/\partial r)F(r,t)$, and $F'' = (\partial/\partial r)^2F(r,t)$. The term $N = N(r,t)$ is defined as \\
\begin{equation}
N(r,t) = \frac{1}{4} (FR + \opensquare F + T) \ ,
\end{equation} \ \\
for notational simplicity.

Note that the gravitational field equation (12)-(14), can be reorganized to yield the following relationships \\
\begin{eqnarray}
\rho = -\ddot{F} +3H^2F +  \frac{b'F}{S^2r^2}  \ , \\
p_r = - 2\dot{H}F + H\dot{F} -3H^2F - \frac{b F}{S^2r^3}  +  \frac{(b'r-b)F'}{2S^2r^2} - \frac{(r-b)F''}{S^2r} \ , \\
p_t = - 2\dot{H}F + H\dot{F} -3H^2F - \frac{(b'r-b)F}{2S^2r^3}  - \frac{(r-b)F'}{S^2r^2} \ ,
\end{eqnarray} \ \\
which can be the generic expressions of the matter threading the wormhole, as a function of the shape function and the specific form of $F(r,t)$. Thus, by specifying the above functions, one deduces the matter content of the wormhole.

one may now adopt several strategies to solve the field equations. For instance, if $b(r)$, and $S(t)$ are specified, and using a specific equation of state $p_r = p_r(\rho)$ or $p_t = p_t(\rho)$ one can obtain $F(r,t)$ from the gravitational field equations and the curvature scalar in a parametric form, $R(r,t)$, from its definition via the metric. Then, once $T = T_{\mu}^{\mu}$ is known as a function of $r,t$, one may in principle obtain $f(R)$ as a function of $R$ from equation (3). For static case, $S(t)=1$, the gravitational field equations (16)-(18) reach to the field equations are obtained by Francisco S. N. Lobo, and  Miguel A. Oliveira [18]. \\

\section{The Wormhole Solutions}
In this section, we obtain  the dynamic wormhole solutions for a specific equation of state.  An interesting equation of state is that of the traceless stress-energy tensor,$T=-\rho+p_r+2p_t=0$, which is usually associated to the Casimir effect, with a massless field. Note that the Casimir effect is sometimes theoretically invoked to provide exotic matter to the system considered at hand. The condition $T=0$ together with equations (16)-(18), provide the following equation \\
\begin{eqnarray}
\fl \ddot{F} + 3H\dot{F} - (12H^2+6\dot{H})F - \frac{2b'F}{S^2r^2} + \frac{(b'r+3b-4r)F'}{2S^2r^2} - \frac{(r-b)F''}{S^2r} =0 .
\end{eqnarray} \ \\
In principle, as mentioned above one may deduce $F(r,t)$ by imposing a specific $b(r)$ and $S(t)$, the specific form $f(R)$ may be found from the trace equation (3). For instance, we consider that the specific  shape function and  scale factor given by \\
\begin{eqnarray}
\eqalign {b(r)=r_0 \\
S(t)=S_0 t^n \ \ , \ \  n > 0 \ ,}
\end{eqnarray} \ \\
where $n$ is a parameter which we take it positive. Thus, equation (19) yields the following solution \\
\begin{eqnarray}
F(t) = c_1 \ t^{m_{+}} + c_2 \ t^{m_{-}} \ , \\ \\
m_{\pm} = -\frac{3n}{2} + \frac{1}{2}  \pm  \frac{1}{2} \sqrt{57n^2-30n+1}
\end{eqnarray} \ \\
$c_1$ and $c_2$ are constant.

The gravitational field equations, (16)-(18), provide  the following relations for the stress-energy tensor threading the wormhole \\
\begin{eqnarray}
\rho =  \frac{27n}{2t^2} \left( c_1a_{+} t^{m_{+}} +  c_2a_{-} t^{m_{-}} \right) \ , \\ \nonumber \\ \nonumber \\
p_r =  \frac{9n}{2t^2} \left[ \left(a_{+} - \frac{2r_0t^2}{9nt^{2n}r^3} \right)c_1 t^{m_{+}} + \left(a_{-} - \frac{2r_0t^2}{9nt^{2n}r^3} \right) c_2 t^{m_{-}} \right] \ , \\ \nonumber \\ \nonumber \\
p_t =  \frac{9n}{2t^2} \left[ \left(a_{+} + \frac{r_0t^2}{9nt^{2n}r^3} \right)c_1 t^{m_{+}} + \left(a_{-} + \frac{r_0t^2}{9nt^{2n}r^3} \right) c_2 t^{m_{-}} \right] \ , \\ \nonumber \\ \nonumber \\ \nonumber
a_{\pm} = - n + \frac{5}{9} \pm \frac{1}{9} \sqrt{57n^2-30n+1} \ .
\end{eqnarray} \ \\
For the specific shape function and scale factor considered above, the Ricci scalar, equation (11), is given by $R = 6n(2n-1)/t^2$. Substituting these relations into the consistency relation (3), provides the specific form of $f(R)$, which is given by \\
\begin{eqnarray}
f(R)= -R \left[ c_1 \left( \frac{6n(2n-1)}{R} \right)^{\frac{m_{+}}{2}}   +   c_2 \left( \frac{6n(2n-1)}{R} \right)^{\frac{m_{-}}{2}} \right] \ .
\end{eqnarray} \ \\

It must be noted that for our solutions we adopted $S(0)=0$ as an initial condition. Spacetimes with accelerated expansion come to play an important role in cosmology. The accelerated expansion occurs for $n > 1$. Note that $n=1$ is the boundary between the accelerated and decelerated expansion. \\

\section{Energy Conditions}
A fundamental point in wormhole physics is the energy condition violations, as mentioned above. However, a subtle issue needs to be pointed out in modified theories of gravity, where the gravitational field equations differ from the classical relativistic Einstein equations. In this section we will check the null energy condition (NEC), and the weak energy condition (WEC) for our solutions in the previous section. The weak energy condition says that the energy density of any system at any point of spacetime for any observer is positive. When the observer moves at the speed of light it has a well defined limit called the null energy condition. The weak and the null energy conditions are the weakest of the energy conditions their violation signals that the other energy conditions  are also violated [19].

The NEC specifies that for any null vector: $T_{\mu\nu}^e k^{\mu}k^{\nu} \geq 0$ (where $k^{\mu}$ is a null vector). The null energy condition leads to the following relationship \\
\begin{equation}
\rho^e + p_r^e = \frac{\rho+p_r}{F} + \frac{\ddot{F}-H\dot{F}}{F} \ .
\end{equation} \ \\
Using the gravitational field equations, (28) takes the familiar form \\
\begin{equation}
\rho^e + p_r^e =  \frac{1}{t^2} \left(2n - \frac{r_0t^2}{t^{2n} r^3}\right) \ .
\end{equation} \ \\
Then, we will have the following inequalities \\
\begin{eqnarray}
\eqalign {2n \geq (t^2r_0)/(t^{2n} r^3) \ \ \longrightarrow \ \  \frac{\rho+p_r}{F} + \frac{\ddot{F}-H\dot{F}}{F} \ \ \geq \ \ 0 \ , \\
2n < (t^2r_0)/(t^{2n} r^3) \ \ \longrightarrow \ \  \frac{\rho+p_r}{F} + \frac{\ddot{F}-H\dot{F}}{F} \ \ < \ \ 0 \ .}
\end{eqnarray} \ \\
We consider that the matter threading the wormhole obeys the energy conditions. The weak energy condition (WEC) is given by $\rho \geq 0$ and $\rho+p_r \geq 0$, then equations (16) and (17), together with (20) yield the following inequalities \\
\begin{eqnarray}
\eqalign {\rho \geq 0 \ \ \longrightarrow \ \  \frac{3n^2}{t^2}F - \ddot{F} \ \ \geq \ \ 0 \ , \\
\rho + p_r \geq 0 \ \ \longrightarrow \ \  -\ddot{F} + \frac{n}{t} \dot{F} + \frac{2n}{t^2} F - \frac{r_0}{r^3 t^{2n}} F \ \ \geq \ \ 0 \ .}
\end{eqnarray} \ \\
Thus, if one imposes that the matter threading the wormhole satisfies the energy conditions, we emphasize that it is the higher derivative curvature terms that sustain the wormhole geometries. Thus, in finding wormhole solutions it is fundamental that the functions $f(R)$ obey inequalities (30) and (31).

Equation (24) yields the following inequalities \\
\begin{eqnarray}
\fl \eqalign {\rho \geq 0 \ \ \longrightarrow  \ \ \cases{c_1 \geq 0 \ , \ c_2 > 0 \ \ or \ \ c_1 > 0 \ , \ c_2 \geq 0 \ \ \ \ \ \ for \ \ \ 0 < n \lesssim 0.04 \ , \\ \\
c_2 \leq 0 \ , \ c_1 > 0 \ \ or \ \ c_2 < 0 \ , \ c_1 \geq 0 \ \ \ \ \ \ for \ \ \ 0.49 \lesssim n < 2 \ , \\ \\
c_1 \leq 0 \ , \ c_2 < 0 \ \ or \ \ c_1 < 0 \ , \ c_2 \leq 0 \ \ \ \ \ \ for \ \ \ n > 2 \ .}}
\end{eqnarray} \ \\
Also, by Using (24) and (25), we reach to the following relationship \\
\begin{equation}
\rho + p_r =  \frac{18n}{t^2} \left[ \left(a_{+} - \frac{r_0t^2}{18nt^{2n}r^3} \right)c_1 t^{m_{+}} + \left(a_{-} - \frac{r_0t^2}{18nt^{2n}r^3} \right) c_2 t^{m_{-}} \right] \ .
\end{equation} \ \\
By inspection we see that the expression (33) can be positive (or zero). \\
For $n > 2$ the stress-energy tensor satisfies the NEC and WEC for the values $c_1 < 0 \ , \ c_2 \leq 0$ or $c_1 \leq 0 \ , \ c_2 < 0$. Also, for $0.49 \lesssim n < 2$ the  stress-energy tensor satisfies the NEC and WEC for the values $c_1=0 \ , \ c_2 < 0$. But, for $0 < n \lesssim 0.04$ the violation of the NEC and WEC can be avoided at some times or in some regions with specific radii. \\
Therefore, we have wormhole spacetimes  with accelerated expansion that the matter threading these spacetimes  satisfies the  energy conditions (NEC and WEC), for the specific values $c_1$ and $c_2$. \\

\section{Conclusion}
In this paper, we have discussed dynamic wormhole solutions of $f(R)$ theory of gravity. The matter needed to support the geometry was assumed to be traceless. In the analysis outlined above, we considered a constant redshift function, which simplified the calculations considerably, yet provide interesting enough exact solutions. Also, we considered a constant shape function  and the scale factor as a positive power of $t$. One may also generalize the results of this paper by considering another suitable shape functions and scale factors.

We have checked the NEC and WEC violations for the solutions. The matter threading the wormholes satisfies the energy conditions (NEC and WEC) for the specific values $c_1$, $c_2$, $n$, and it is the higher order curvature derivative terms, that may be interpreted as a gravitational fluid, that support these wormholes, fundamentally different from their counterparts in general relativity.

In this work, we have shown that the matter threading the wormhole spacetimes  with either accelerated expansion or  decelerated expansion satisfies the NEC and WEC for the specific values $c_1$ and $c_2$. \\


\section*{References}
\begin{thebibliography}{19}

\bibitem {1} Spergel D N  \emph{et al.} {\it Preprint} arXiv:astro-ph/0603449.


\bibitem {2} Astier P  \emph{et al.} 2006 {\it Astron. Astrophys.} {\bf 447}, 31


\bibitem {3} Weinberg S  1989 {\it Modern Phys. Rev.} {\bf 61}, 1


\bibitem {4} Carroll S M,  Press W H and Turner E L 1992 {\it Ann. Rev. Astron. Astrophys.} {\bf 30}, 499


\bibitem {5} Starobinsky A A 1980 {\it Phys. Lett.} B {\bf 91}, 99


\bibitem {6} Carroll S M, Duvvuri V, Trodden M  and Turner M S  2004 {\it Phys. Rev.} D {\bf 70}, 043528


\bibitem {7} Morris M S and  Thorne K S 1988 {\it Am. j. Phys.} {\bf 56}, 395


\bibitem {8} Visser M 1995 {\it Lorentzian Wormholes: from Einstein to
Hawking} (AIP Press, Singapore)


\bibitem {9} Hochberg D and Visser M   {\it Preprint} arXiv:gr-qc/9901020.


\bibitem {10} Visser M and Hochberg D {\it Generic wormhole throat.}  {\it Preprint} arXiv:gr-qc/9710001.


\bibitem {11} Hochberg D and Visser M  1998 {\it Phys. Rev. Lett.}  {\bf 81}, 746


\bibitem {12} Hochberg D and Visser M  1998  {\it Phys. Rev.} D {\bf 58}, 044021


\bibitem {13} Bhawal B and Kar S 1992 {\it Phys. Rev.} D {\bf 46}, 2464


\bibitem {14} Das A  2003  {\it Nucl. Phys.} B {\bf 653}, 279


\bibitem {15} Bronnikov K A and  Kim S W 2003 {\it Phys. Rev.} D {\bf 67}, 0640027


\bibitem {16} Capozziello S 2002 {\it Int. J. Mod. Phys.}  {\bf 11}, 483


\bibitem {17} Roman T A 1993 {\it Phys. Rev.} D {\bf 47}, 1370


\bibitem {18} Lobo F S N  and Oliveira M A 2009 {\it Phys. Rev.} D {\bf 80}, 104012


\bibitem {19} Hawking S W  and Ellis G F R 1973 {\it The Large Scale
Structure of Spacetime} (Cambridge University Press, Cambridge)

\end{thebibliography}
\end{document}